# An Active Inference Model of Collective Intelligence

**Rafael Kaufmann[1], Pranav Gupta[2], and Jacob Taylor [3,4]\***

[1] Independent researcher
[2] Tepper School of Business, Carnegie Mellon University
[3] Institute of Cognitive & Evolutionary Anthropology, University of Oxford
[4] Crawford School of Public Policy, Australian National University
\* Correspondence: jacob.taylor@anthro.ox.ac.uk

**Abstract:** Collective intelligence, an emergent phenomenon in which a composite system of multiple interacting agents performs at levels greater than the sum of its parts, has long compelled research efforts in social and behavioral sciences. To date, however, formal models of collective intelligence have lacked a plausible mathematical description of the relationship between local-scale interactions between highly autonomous sub-system components (individuals) and global-scale behavior of the composite system (the collective). In this paper we use the Active Inference Formulation (AIF), a framework for explaining the behavior of any non-equilibrium steady state system at any scale, to posit a minimal agent-based model that simulates the relationship between local individual-level interaction and collective intelligence (operationalized as system-level performance). We explore the effects of providing baseline AIF agents (Model 1) with specific cognitive capabilities: Theory of Mind (Model 2); Goal Alignment (Model 3), and Theory of Mind with Goal Alignment (Model 4). These stepwise transitions in sophistication of cognitive ability are motivated by the types of advancements plausibly required for an AIF agent to persist and flourish in an environment populated by other AIF agents, and have also recently been shown to map naturally to canonical steps in human cognitive ability. Illustrative results show that stepwise cognitive transitions increase system performance by providing complementary mechanisms for alignment between agents' local and global optima. Alignment emerges endogenously from the dynamics of interacting AIF agents themselves, rather than being imposed exogenously by incentives to agents' behaviors (contra existing computational models of collective intelligence) or top-down priors for collective behavior (contra existing multiscale simulations of AIF). These results shed light on the types of generic information-theoretic patterns conducive to collective intelligence in human and other complex adaptive systems.

## 1. Introduction

Subjectively, we perceive ourselves to be autonomous individuals at the same time that we actively participate in collectives. Families, organizations, sports teams, and polities exert agency over our individual behavior [1,2] and are even capable, under certain conditions, of intelligence that cannot be explained by aggregation of individual intelligence [3,4]. To date, however, formal models of collective intelligence have lacked a plausible mathematical description of the functional relationship between individual and collective behavior.

In this paper, we use the Active Inference Formulation (AIF), a process theory of the Free Energy Principle (FEP), to develop a clearer understanding of the relationship between patterns of individual interaction and collective intelligence. We adopt a definition of collective intelligence established within organizational psychology, as groups of individuals capable of acting collectively in ways that seem intelligent and that cannot be explained by individual intelligence [5] (p.3) [3]. In more general terms, collective intelligence is an emergent phenomenon in which a composite system of multiple interacting agents performs at levels greater than the sum of its parts. We have a particular interest in elucidating mechanisms and dynamics that explicate human collective intelligence, but the universality of our formal computational approach makes it relevant to collective intelligence in any complex adaptive system.

We suggest that a formal account of collective intelligence must crucially entail a testable framework for explaining relationships between behavior across multiple scales—e.g., between individual interaction and collective performance. Existing accounts of collective intelligence, particularly those focused on human collective intelligence, are marred precisely by a lack of alignment between individual and collective scales of analysis. Accounts of local-scale interactions tend to construe individuals as 1st-person, goal-directed agents endowed with discrete cognitive mechanisms (specifically social perceptiveness or Theory of Mind and shared intentionality; see [6,7]) that allow individuals to establish and maintain adaptive connections with other individuals in service of shared goals [3–5,8–10].[1] Researchers conjecture that these mechanisms allow collectives to derive and utilize more performance-relevant information from the environment than could be derived by an aggregation of the same individuals acting without such connections (for example, by facilitating an adaptive, system-wide balance between cognitive efficiency and diversity; see [4]). Empirical substantiation of such claims has proven difficult, however. Most investigations rely heavily on laboratory-derived summaries or "snapshots" of individual and collective behavior that flatten the complexity of local scale interactions [11] and make it difficult to examine causal relationships between individual scale mechanisms and collective behavior as they typically unfold in real world settings [12,13].

Accounts of global-scale (collective) behavior, by contrast, tend to adopt system-based (rather than agent-based) perspectives that render collectives as random dynamical systems in phase space [14–17]. Only rarely deployed to assess the construct of human collective intelligence specifically (e.g., [18]), these approaches have been fruitful for identifying the types of phase-space dynamics (such as synchrony, metastability, or symmetry breaking) that correlate with collective performance more generally construed [19–23]. However, on their own, such analyses are limited in their ability to generate testable predictions for multiscale behavior, such as how global-scale dynamics (rendered in phase-space) translate to specific local-scale interactions (in state-space), or how local-scale interactions between individuals translate to evolution and change in global-scale dynamics [17].

In sum, the substantive differences between these two analytical perspectives (individual and collective) on collective intelligence make it difficult to formulate a formal description of how details of local-scale interactions between individuals relate to global-scale collective behavior and vice versa. Most urgent for the development of a formal model of collective intelligence, therefore, is a common mathematical framework capable of operating between individual-level cognitive mechanisms and system-level dynamics of the collective [4].

---

[1] Reidl and colleagues [10] report a recent analysis of 1356 groups that found social perceptiveness and group interaction processes to be strong predictors of collective intelligence measured by a psychometric test.

*1.1 The Free Energy Principle and an Active Inference Formulation of Collective Intelligence*

FEP has recently emerged as a candidate for this type of common mathematical framework for multiscale behavioral processes [24–26]. FEP states that any non-equilibrium steady state system self organizes as such by minimizing variational free energy in its exchanges with the environment [27]. The key trick of FEP is that the principle of free energy minimization can be neatly translated into an agent-based process theory, AIF, of approximate Bayesian inference [28] and applied to any self-organizing biological system at any scale [29]. The upshot is that, in theory, any AIF agent at one spatio-temporal scale could be simultaneously composed of nested AIF agents at the scale below, and a constituent of a larger AIF agent at the scale above it [30–32]. In effect, AIF allows you to pick a composite agent A that you want to understand, and it will be generally true both that: A is an approximate, global minimizer of free energy at the scale at which that agent reliably persists; and A is composed of subsystems {A_i} that are approximate, local minimizers of free energy (which is composed of the remainder of A). Thus, supposing that human individuals and the collectives they constitute both minimally satisfy the necessary assumptions of FEP [pertaining to (a) ergodic and (b) non equilibrium steady state systems; see 24], collective intelligence could conceivably be modelled as a case of individual AIF agents that interact within—or indeed, interact to produce—a superordinate AIF agent at the scale of the collective [33,34]. AIF thus provides a framework within which an agent-based model of collective intelligence could be developed.

An AIF model of collective intelligence begins with the depiction of a minimal AIF agent. Specifically, an AIF agent denotes any set of states enclosed by a "Markov blanket"—a statistical partition between a system's internal states and external states [35]—that infers beliefs about the causes of (hidden) external states by developing a probabilistic *generative model* of external states [27]. A Markov blanket is composed of sensory states and active states that mediate the relationship between a system's internal states and external states: external states ($\psi$) act on sensory states ($s$), which influence, but are not influenced by internal states ($b$). Internal states couple back through active states ($a$), which influence but are not influenced by external states. Through conjugated repertoires of perception and action, the agent embodies and refines (learns) a generative model of its environment [36] and the environment embodies and refines its model of the agent (akin to a circular process of environmental niche construction; see [37]).

Having established the notion of an agent, the next step is to consider the existence of multiple nested AIF agents across multiple scales of behavioral organization, beginning with individual and collective scales. Existing multiscale treatments of AIF provide a clear account of "downward reaching" causation of multiscale biological systems, whereby superordinate AIF agents systematically determine (or "enslave;" see [38]) the behavior of subordinate AIF agents [30,33,39]. For instance, while cells or neurons could be modelled as free energy minimizing agents in their own right, the behavioral degrees of freedom available to these individual agents as members of a superordinate entity (a multicellular organism or a brain) are fundamentally constrained by the dynamics of the superordinate entity [40].

While useful for depicting the behavior of neurons within brains or cells within multicellular organisms, this general account of multiscale AIF is yet to be specified for modelling other behaviors of interest, including intelligent behavior of human collectives [34,41]. Human interaction unfolds between highly autonomous individuals whose statistical boundaries for self-evidencing are often transient, distributed, and multiple [42–45]. Unlike cells or neurons, which rely on simple autoregulatory mechanisms to sustain participation in collective ensembles, human agents participate in collectives by leveraging an array of phylogenetic (evolutionarily) and ontogenetic (developmental) mechanisms, and socio-culturally constructed regularities or affordances [37,46,47]. Human agents' cognitive abilities and sociocultural niches create avenues for active participation in functional collective behavior (e.g., the pursuit of shared goals), as well as avenues to shirk global constraints in the pursuit of local (individual) goals.

Existing toy models that simulate the emergence of collective behavior under AIF do so by simply using the statistical constraints from one scale to drive behavior at another, either explicitly(e.g., by endowing AIF agents with a genetic prior for functional specialization within a superordinate system; see [33]) or by constructing a scenario in which a superordinate agent (or shared generative model) is predestined by virtue of the fact that agents' environment is constituted only by the sensory generated by each other [48,49]. While these scenarios successfully depict the formation and persistence of collectives (as a function of local-scale interactions between individual AIF agents), and while that collective may be capable of intelligence, these dual characteristics depend on the type and quality of instructions supplied exogenously (or from the "top-down"), more so than from the basic information-theoretic patterns of complex adaptive systems implied by AIF (and emergent from the "bottom-up"). In this sense, extant models of AIF bear a closer resemblance to Searle's [50] "Chinese Room Argument" than to what we would recognize as emergent collective intelligence.

Currently missing from models of multiscale AIF are specifications for how a system's emergent cognitive capabilities causally relate to individual agents' emergent cognitive capabilities, and thus how local-scale interactions between individual AIF agents give rise, *endogenously*, to superordinate AIF agents that exhibit behaviors of interest [34]. Specifically, existing approaches lack a description of the key cognitive mechanisms that might provide a mechanistic "missing link" for collective intelligence under AIF. The key puzzle of collective intelligence that we attempt to explain is when and how intelligent behavior at the individual agent level gives rise to intelligent behavior at the collective level. Thus, in this paper we ask: what basic information theoretic patterns of interaction between individual AIF agents create opportunities for collective intelligence at the global scale?

*1.2 Our approach*

To operationalize AIF in a way that is useful for investigating this question, we begin by examining what minimal features of individual AIF agents are required to achieve collective intelligence. We conjecture that very generic information theoretic patterns of an environment in which individual AIF agents exploit other AIF agents as affordances of free energy minimization should support the emergence of collective intelligence, operationalized as active inference (or free energy minimization) at the level of the global-scale system. Importantly, we expect that these patterns emerge under very general assumptions and from the dynamics of AIF itself—without the need for exogenously imposed fitness or incentive structures on local-scale behavior, contra extant computational models of collective intelligence (that rely on cost or utility functions; e.g., Reia et al., 2019; Krafft, 2019) or other common approaches to reinforcement learning (that rely on exogenous parameters of the Bellman equation; see [51,52]).

To justify our modelling approach, we draw upon recent research that systematically maps the complex adaptive learning process of AIF agents to empirical social scientific evidence for cognitive mechanisms that support adaptive human social behavior. We posit a series of stepwise progressions or "hops" in the individual cognitive ability of any AIF agent in an environment populated by other self-similar AIF agents. These hops represent evolutionarily plausible "adaptive priors" [32] (p.109) that would likely guide action-perception cycles of AIF agents in a collective toward unsurprising states:

- **Baseline AIF -** AIF agents, to persist as such, will minimize immediate free energy by accurately sensing and acting on salient affordances of the environment. This will require a general ability for "perceptiveness" of the (physical) environment.

- **Folk Psychology -** AIF agents in an environment populated by other AIF agents would fare better by minimizing free energy not only relative to their physical environment, but also to the "social environment" composed of their peers [46]. The most parsimonious way for AIF agents to derive information from other agents would be to (i) assume that other agents are self-similar, or are "creatures like me" [53], and (ii) differentiate other-generated information by calculating how it diverges from self-generated information (akin to a process of "alterity" or self-other distinction). This ability aligns with the notion of a "folk psychological theory of society," in which humans deploy a combination of phylogenetic and ontogenetic modules to process social information [54,55].

- **Theory of Mind -** AIF agents that develop "social perceptiveness" or an ability to accurately infer beliefs and intentions of other agents will likely outperform agents with less social perceptiveness. Social perceptiveness, also commonly known in cognitive psychology as "Theory of Mind," would minimally require cognitive architecture for encoding the internal belief states of other agents as a source of self-inference (for game-theoretical simulations of this proposal, see [56,57]). As discussed above, experimental evidence suggests that social perceptiveness or Theory of Mind (measured using the "Reading the Mind in the Eyes" test; see [58]) is a significant predictor of human collective intelligence in a range of in-person and on-line collaborative tasks [4].

- **Goal Alignment -** It is possible to imagine scenarios in which the effectiveness of Theory of Mind would be limited, such as situations of high informational uncertainty (in which other agents hold multiple or unclear goals), or in environments populated by more agents than would be computationally tractable for a single AIF agent to actively theorize [59]. AIF agents capable of transitioning from merely encoding internal belief states of other AIF agents to recognizing shared goals and actively aligning goals with other AIF agents would likely enjoy considerable coordination benefits and (computational) efficiencies [7,60] that would also likely translate to collective-level performance [61,62].

- **Shared Norms -** Acquisition of capacities to engage directly with the reified signal of sharedness (a.k.a., "norms") between agents as a stand-in for (or in addition to) bottom-up discovery of mutually viable shared goals would also likely confer efficiencies to individuals and collectives [37]. Humans appear unique in their ability to leverage densely packaged socio-cultural installed affordances to cue regimes of perception and action that establish and stabilize adaptive collective behavior (a process recently described as "Thinking Through Other Minds"; see [47]).

The clear resonance between generic information-theoretic patterns of basic AIF agents and empirical evidence of human social behavior is remarkable, and gives credence to the extension of seemingly human-specific notions such as "alterity", "shared goals", "alignment", "intention," and "meaning" to a wider spectrum of bio-cognitive agents [63]. In effect, the universality of FEP—a principle that can be applied to any biological system at any scale—makes it possible to strip-down the complex and emergent behavioral phenomenon of collective intelligence to basic operating mechanisms, and to clearly inspect how local-scale capabilities of individual AIF agents might enable global-scale state optimization of a composite system.

In the following section we use AIF to model the relationship between a selection of these hops in cognitive ability and collective intelligence. We construct a simple 1D search task [based on 64], in which two AIF agents interact as they pursue individual and shared goals. We endow AIF agents with two key cognitive abilities—Theory of Mind and Goal Alignment—and vary these abilities systematically in four simulations that follow a 2x2 (Theory of Mind x Goal Alignment) progression: Model 1 (Baseline AIF, no social interaction), Model 2 (Theory of Mind without Goal Alignment), Model 3 (Goal Alignment without Theory of Mind), and Model 4 (Theory of Mind with Goal Alignment). We use a measure of free energy to operationalize performance at the local (individual) and global (collective) scales of the system [65]. While our goals in this paper are exploratory (these models and simulations are designed to be generative, not to test hypotheses), we do generally expect that increases in sophistication of cognitive abilities at the level of individual agents will correspond with an increase in local- and global-scale performance. Indeed, illustrative results of model simulations (Section 3) show that each hop in cognitive ability improves global system performance, particularly in cases of alignment between local and global optima.

## 2. Materials and Methods
### 2.1 Paradigm and set-up

Our AIF model builds upon the work of McGregor and colleagues, who develop a minimal AIF agent that behaves in a discrete one-dimensional time world [64]. In this set-up, a single agent senses a chemical concentration in the environment and acts on the environment by moving one of two ways until it arrives at its desired state, the position in which it believes the chemical concentration to be highest, denoting a food source. We adapt this paradigm by modelling two AIF agents (Agent A and Agent B) that occupy the same world and interact according to parameters described below (see Figure 1). The McGregor et al. paradigm and AIF model is attractive for its computational implementability and tractability as a simple AIF agent with minimum viable complexity. It is also accessible and reproducible; whereas most existing agent-based implementations of AIF are implemented in MATLAB, using the SPM codebase (e.g., [52]), an implementation of the McGregor et al. AIF model is widely available in the open-source programming language Python, using only standard open source numerical computing libraries [66].

We extend the work of McGregor and colleagues to allow for interactions not only between an agent and the "physical" environment, but also between an agent and its "social" environment (i.e., its partner). Accordingly, we make minor simplifications to the McGregor et al. model that are intended to reduce the number of independent parameters and make interpretation of phenomena more straightforward (alterations to the McGregor et al. model are noted throughout).

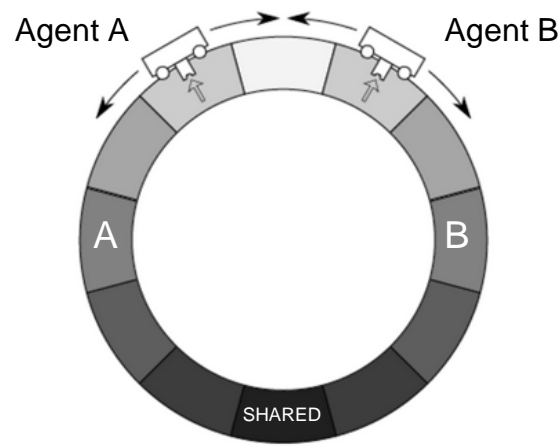

**Figure 1**. A minimal collective system of two AIF agents (adapted from McGregor et al.). We implement two agents (Agent A and Agent B) that have one common target position (Shared Target) and one individual target position (A's Target; B's Target). All targets are encoded with equal desirability. This figure is notional: our simulation environment contains 60 cells instead of the 12 depicted here. Note: we randomize the location of the shared target while preserving relative distances to unshared targets to ensure that the agents' behavior is not an artefact of its location in the sensory environment.

*2.2 Conceptual outline of AIF model*

Our model consists of two agents. Descriptively, one can think of these as simple automata, each inhabiting a discrete "cell" in a one-dimensional circular environment where there are predefined targets (food sources). As agents aren't endowed with a frame of reference, an agent's main cognitive challenge is to situate itself in the environment (i.e., to infer its own position). Both agents have the following capabilities:

- **Physical capabilities:**
    - "Chemical sensors" able to pick up a 1-bit chemical signal from the food source at each time step;
    - "Actuators" that allow agents to "move" one cell at each time step;
    - "Position and motion sensors" that allow agents to detect each other's position and motion.

- **Cognitive capabilities:**
    - Beliefs about their own current position; we construe this as a "self-actualization loop" or Sense->Understand->Act cycle: (1) sense environment; (2) optimize belief distribution relative to sensory inputs (by minimizing free energy given by an adequate generative model); and (3) act to reduce FE relative to desired beliefs, under the same generative model.
    - Desires (also described as "desired beliefs") about their own position relative to their prescribed target positions;
    - Ability to select the actions that will best "satisfy" their desires;
    - "Theory of Mind": they possess beliefs about their partner's position, knowledge of their partner's desires, and therefore, the ability to imagine the actions that their partners are expected to take. We implement this as a "partner-actualization loop" that is formally identical to the self-actualization loop above;
    - "Goal Alignment": the ability to alter their own desires to make them more compatible with their partner's.

*2.3 Model preliminaries*

Throughout, we use the following shorthand:

- $q^{superscript} \triangleq softmax(b^{superscript})$ for any superscript index, where $softmax(b)_i \triangleq \frac{e^{b_i - min(b)}}{\Sigma\, e^{b_i - min(b)}}$. This converts a belief represented as a vector in $\mathbb{R}^N$ to the equivalent probability distribution over [1..N]. $b^{superscript} = \ln q^{superscript} + K$ converts back.
- Beliefs are implicitly constrained to the range $b^i \in \mathbf{B} = [-10, 0]$.
- $\varphi \triangleq \psi^{partner}$ when necessary, to disambiguate between it and $\psi^{own}$.
- $(v_{+x})_i \triangleq v_{i+x}$ to denote shifting a vector.
- $\Theta_\alpha(q) \triangleq \alpha\, q + \frac{1-\alpha}{N}$ to denote "re-ranging" a probability distribution, squishing its range from [0, 1] to $[\frac{1-\alpha}{N}, \alpha - \frac{1-\alpha}{N}]$.
- All arithmetic in the space of positions and actions $(\psi, a, \Delta)$ is considered to be mod N.

*2.4 State space*

These capabilities are implemented as follows. Each agent $A^i$ is represented by a tuple $A^i = (\psi^i, s^i, b^i, a^i)$. In what follows we'll omit the indices except where there is a relevant difference between agents. These tuples form the relevant state space (see Figure 2):

- $\psi \in [0..N-1]$ is the agent's external state, its position in a circular environment with period N. Crucially, the agent doesn't have direct access to its external state, but only to limited information about the environment afforded through the sensory state below.

- $s = (s^{own} \in \{0, 1\}, \Delta \in [0..N-1], a^{pp} \in \{-1, 0, 1\})$ is the agent's sensory state. $s^{own}$ is a one-bit sensory input from the environment; $\Delta$ is the perceived difference between the agent's own position and its partner's; $a^{pp}$ is the partner's last action.

- $b = (b^{own} \in \mathbf{B}^N, b^{*own} \in \mathbf{B}^N, b^{partner} \in \mathbf{B}^N, b^{*partner} \in \mathbf{B}^N)$ is the agent's internal or "belief" state. $b^{own}$ and $b^{*own}$ are, respectively, its actual and desired beliefs about its own position; equivalently, $b^{partner}$ and $b^{*partner}$ are its actual and desired beliefs about its partner's position.

- $a = (a^{own} \in \{-1, 0, 1\}, a^{partner} \in \{-1, 0, 1\})$ is the partner's action state: $a^{own}$ is its own action; $a^{partner}$ is the action it expects from the partner.

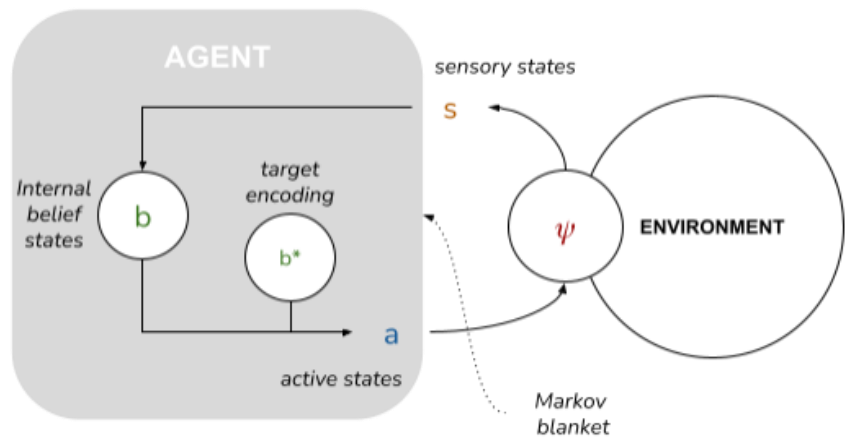

**Figure 2.** AIF agent based on McGregor et al. 2015. A Markov blanket defines conditional independencies between a set of internal belief states ($b$) and a set of environment states ($\psi$) with target encoding or "desires" ($b^*$).

*2.5 Agent Evolution*

These states evolve according to a discrete-time free energy minimization procedure, extended from McGregor et al. (Figure 3).

```
procedure simulate(ψ, b, b∗)
    loop
        s^own i ← random value using P^i(s^own | ψ)
```

```
                    aⁱ ← argminₐ   Fⁱ(b*, b, sⁱ, a)

                    b'ⁱ ← optimise(bⁱ, sⁱ, aⁱ)

                    ψ' ← ψ + a           ◁ Unlike McGregor et al., we assume agents always act as intended

                    Δ ← (ψ¹ - ψ², ψ² - ψ¹)     ◁ Agents' "position sensors" are assumed to be perfect

                    aᵖᵖ ← (ψ'² - ψ², ψ'¹ - ψ¹)  ◁ Agents' "motion sensors" are assumed to be perfect

              end loop
      end procedure

      function optimize(b, s, a)
              b' = b                     ◁ starting from previous epoch's belief
              for i ∈ {1 · · · k} do
                      b' ← b' − η ·∂/∂b' F(b', b, s, a)      ◁ gradient descent on b'
              end for
              return b'
      end function
```

**Figure 3.** Pseudo code for agent evolution (adapted from McGregor et al., 2015). Note that the loop is run for both agents in lockstep, but each agent selects actions and optimizes beliefs individually.

*2.6 Sensory model*

Let us recapitulate McGregor et. al's definition of the free energy for a single-agent model:

$$F(b', b, s, a) = D_{KL}(q(\psi' \mid b') \parallel p(\psi', s \mid b, a)) \tag{1}$$

where $q(b)$ = softmax($b$) is the "variational (probability) density" encoded by $b$, and $p(\psi', s \mid b, a)$ is the "generative (probability) density" representing the agent's mental model of the world (Friston et al. 2006). $D_{KL}$ is the Kullback–Leibler (KL) divergence or relative entropy between the variational and generative densities (see Friston et al., 2009). Evidently, as b is an arbitrary real vector, the optimal b' is the one which produces $q = p$, that is, $b_i = -\log(p_i) + K$.

In order to respect the causal relationships prescribed by the Markov blanket (see Figure 2), the generative density may be decomposed as:

$$p(', s \mid b, a) = P(\psi' \mid s, b, a, \psi) \, P(s \mid b, a, \psi) \, P(\psi \mid b, a) \tag{2}$$

where the three terms within the summation are arbitrary functions of their variables. In the single-agent model, where the only source of information is the environment, we follow McGregor's model, in a slightly simplified form:

1. $P(\psi' \mid s, b, a, \psi) = \delta(\psi', \psi + a)$: the agent's actions are always assumed to have the intended effect, $\delta$ being the discrete Kronecker delta.

2. $P(s \mid \psi) = k^s (1-k)^{1-s} e^{-\omega |\psi - \psi_{mid}|}$: the agent assumes the probability of s = 1 (sensoria triggered) is higher for regions near the "center" of the environment. This is identical to the real "physical" probability of chemical signals, meaning the agent's generative distribution is correct.
3. $P(\psi \mid b, a) = q(b)$, in agreement with the definition of $b$ as encoding the belief distribution over ψ.

From list item 1 directly above, this generative density can also be read as a simple Bayesian updating plus a change of indexes to reflect the effects of the action: $p(\psi', s \mid b, a) = P(s \mid \psi' - a) P(\psi' - a \mid b)$ or even more simply, $p_{\psi'}^{posterior} = p_{\psi'-a}^{s} p_{\psi'-a}^{prior}$.

In our model, both agents implement their own copies of the generative density above (we leave it to the reader to add "☐$^{own}$" indices where appropriate). The parameter $k$, denoting the maximum sensory probability, is assumed agent-specific; we naturally identify it with an agent's "perceptiveness". ω and $\psi_0$, on the other hand, are environmental parameters.

*2.7 Partner model*

In addition to the sensory model, we will define a new generative density implementing the agent's inference of its partner's behavior, or "Theory of Mind" (ToM; see Figure X). An agent with a sensory and partner model will adopt the following form:

p(', Δ, $a^{pp}$ | b, a) = P(' | Δ, $a^{pp}$, b, $a^{partner}$, ) P(Δ | b, $a^{partner}$, ) P($a^{pp}$ | b, $a^{partner}$, ) P( | b, $a^{partner}$)     (3)

Where:
1. $P(\varphi' \mid \Delta, a^{pp}, b, a^{partner}, \varphi) = \delta(\varphi', \varphi + a^{partner})$: the partner's actions are always assumed to have the intended effect.
2. $P(\Delta \mid b, a^{partner}, \varphi) = P(\psi = \varphi + \Delta \mid b, \varphi) = q_{-\Delta}^{own}$; the probability of observing a given Δ, given the partner's position φ, is equal to the probability the agent ascribes to itself being in the corresponding position $\psi = \varphi + \Delta$. This holds as we assume Δ is a deterministic variable of the external states.
3. $P(a^{pp} \mid b, a^{partner}, \varphi) = P(a^{partner} \mid \varphi - a^{pp}, b^{\star partner})$, where

$$P(a^{partner} = 0 \mid \varphi) = \xi \frac{q^{\star partner}}{max(q^{\star partner})}$$

$$P(a^{partner} = \pm 1 \mid \varphi) = \left\{1 - \xi \frac{q^{\star partner}}{max(q^{\star partner})}\right\} \frac{1}{q_{-1}^{\star partner} + q_{+1}^{\star partner}} q_{-a^{partner}}^{\star partner}$$

4. $P(\varphi \mid b, a^{partner}) = q^{partner}$ defines the "prior" and is analogous to (3) in the sensory model.

The unwieldy formula in list item 3 above deserves some additional discussion. Unlike the sensory model, where we can assume the agent "knows" the correct probability distribution for *s*, here the agent must produce probabilities of the partner's actions in abstract, whereas the "true" probabilities are only truly known in the partner's mind (i.e., given the partner's *actual* internal states). To do so, we provide the agent with a simple mechanistic model of the partner's actions.

Remember the agent is assumed to know the partner's desires, $q^{\star partner}$. We correspondingly assume the partner will act according to those desires, i.e., the higher a partner's desire for its current location, the more likely it is to stay put. In order to eliminate spurious dependence on absolute values of $q^{\star partner}$, we set $P(a^{partner} = 0)$ to be proportional to $q^{*partner}/max(q^{*partner})$, fixing the maximum probability at $\xi$ (when $q^{*partner}$ achieves its global maxima). This leaves the remainder $\xi \frac{q^{*partner}}{max(q^{*partner})}$ to be allocated across the other actions ($\pm 1$), which we do by assuming the probability of moving in a given direction is proportional to the desires in the adjacent locations (see Figure 4 below).

Although these formulas seem complicated, especially list item 3 above, they result in a generative density has the same form as the original generative density from the baseline sensory model, $p_{\varphi'}^{posterior} = p_{\varphi'-a^{partner}}^{\Delta,a^{pp}} p_{\varphi'-a^{partner}}^{prior}$. This is consistent with our modeling decision to make the "other-evidencing loop" functionally identical to the "self—actualization loop", as discussed above (Section 2.2).

As before, each agent implements its own copy of the partner model. $\xi$ is assumed equal for both agents; they have the same capability to interpret the partner's actions.

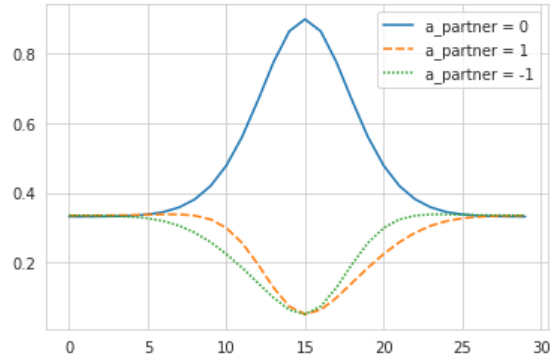

**Figure 4.**   Illustrative plot of $P(a^{partner})$.

*2.8 Agent-level free energy*

We are finally ready to define the free energy for our individual-level model. For each agent:

$$F = D_{KL}(q'^{own} \parallel p^{own} \, \Theta_\alpha(p_{+\Delta'}^{partner})) + D_{KL}(q'^{partner} \parallel p^{partner} \, \Theta_{\alpha^2}(p_{-\Delta'}^{own})) \tag{4}$$

Where:
1.  $p_{\psi'}^{own} = P(s^{own} \mid \psi' - a^{own}) \, q_{\psi'-a^{own}}^{own}$ is the sensory model (outlined above in Section 2.6).
2.  $p_{\varphi'}^{partner} = q_{\varphi'-a^{partner}+\Delta}^{own} P(a^{partner} \mid \varphi' - a^{partner} - a^{pp}, b^{*partner}) \, q_{\varphi'-a^{partner}}^{partner}$ is the partner model (outlined above in subsection 2.7).
3.  The "reranging" function, $\Theta_\alpha$, serves to moderate the influence of the partner model on the agent's own beliefs, and vice-versa. $\alpha$ is an agent-specific parameter, which, as we will see in Subsection 2.9, is identified with each agent's degree of "alterity."
4.  The right-hand side of each KL divergence (i.e., the products of generative densities) is implicitly constrained to $[e^{-10}, 1]$, to ensure the resulting beliefs remain within their range $B$. This is interpreted as preventing overconfidence and is implemented as a simple maximum.

We interpret eq. 4 as follows: The agent's sensory and partner models jointly constrain its beliefs both about its own position and its partner's position. Thus, at each step, the agent: (a) refines its beliefs about both positions, in order to best fit the evidence provided by all of its inputs (i.e., its "chemical" sensor for the physical environment and "position and motion" sensor for its partner); and (b) selects the "best" next pair of actions (for self and partner), i.e., that which minimizes the "difference" (the KL divergence) between its present beliefs and the desired beliefs.[2]

### 2.9 Theory of Mind

In this section we motivate the parameterization of an agent's Theory of Mind ability with $\alpha$, or simply, its degree of *alterity*.

Note that when considered as a discrete-time dynamical system evolution, the process of refining beliefs about own and partner positions in the environment (step (a) in Section 2.8 above) potentially involves multiple recursive dependencies: the updated variational densities $q'^{own}$ and $q'^{partner}$ both depend on the previous $q^{own}$ (via both $p^{own}$ and $p^{partner}$), as well as on the previous $q^{partner}$ (via $p^{partner}$). This is by design: the dependencies ensure that $q'^{own}$ and $q'^{partner}$ are consistent with each other, as well as with their counterparts across time steps. However, too much of a good thing can be a problem. If left unconstrained, $q'^{own}$ and $q'^{partner}$ can easily evolve towards spurious fixed points (Kronecker deltas), which can be interpreted as overfitting on prematurely established priors.[3] On the other hand, if $q'^{own}$ were to depend only on $q^{own}$, it would eliminate the spurious fixed points: without the crossed dependence, the first term of the partner model (Section 2.7) only has fixed points at $(q'^{own} = \delta(\psi', argmax(q^{\star\,own})), a^{own} = 0)$, meaning that the agent has achieved a local desire optimum. Effectively, this "shuts down" the agent's ability to use the partner's information to shape its own beliefs, or its theory of mind, making it equivalent to MacGregor's original model.

Thus, there would appear to be no universal "best" value for an agent's Theory of Mind; an appropriate level of Theory of Mind would depend on a trade-off between the risk of overfitting and that of discarding valid evidence from the partner. The appropriate level of Theory of Mind would also depend on the agent's other capabilities (in this case, its perceptiveness, *k*).

This motivates the operationalization of $\alpha$ as a parameter for the intensity to which Theory of Mind shapes the agent's beliefs. $\alpha$ can be understood simply as an agent's degree of *alterity*, or propensity to see the "other" as an agent like itself. In simulations with values of $\alpha$ close to 0, we expect the partner's behavior to be dominated by its own "chemical" sensory input. Increasing $\alpha$, we expect to see an agent's behavior being more heavily influenced by inputs from its partner, driving $q^{own}$ to become sharper as soon as $q^{partner}$ does so. Past a certain threshold, this could spill over into premature overfitting.

---

[2] For reasons of numerical stability, we follow McGregor et al. in implementing (b) before (a): The agent chooses the next actions based on current beliefs, then updates beliefs for the next time-step, based on the expected effects of those actions [64] (pp.6-7)

[3] In this case, it could be possible to observe scenarios such as *"the blind leading the blind"* in which a weak agent fixates on the movement trajectory of a strong agent who is overconfident about its final destination.

Finally, note the $\alpha^2$ in the second term of agent-level free energy (eq. 4). This represents the notion that the agent is using "second-order theory of mind" or thinking about what its partner might be thinking about it.[4] Here, $p^{own}$ comes in as "my model of my partner's model of my behavior". It seems appropriate for the agent to believe the partner to possess the same level of alterity as itself; we then represent this as applying the rearranging function (the "squishing" of the probability distribution) twice, $\Theta_\alpha \bullet \Theta_\alpha = \Theta_{\alpha^2}$.

*2.10 Goal Alignment*

In this section we motivate the parameterization of the degree of goal alignment between agents.

Recall that $b^{\star\,own}$ is an arbitrary (exogenous) real vector; the implied desire distribution can have multiple maxima, leading to a generally challenging optimization task for the agent. Theory of Mind can help, but it can also make matters worse: if $b^{\star\,partner}$ also has multiple peaks, the partner's behavior can easily become *ambiguous,* i.e., it could appear coherent with multiple distinct positions. This ambiguity can easily lead the agent astray.

This problem is reduced if the agents have the ability to *align goals* with each other, that is, to avoid pursuing targets that are not shared between them. We implement this as:

$$b^{\star\,own} \leftarrow b^{\star\,shared} + (1-\gamma)b^{\star\,own}_{private} \quad (5)$$

$$b^{\star\,partner} \leftarrow b^{\star\,shared} + (1-\gamma)b^{\star\,partner}_{private} \quad (6)$$

where $\gamma$ is a parameter representing the degree of alignment between this specific agent pair, and we assume each agent has knowledge of what goals are shared vs private to itself or its partner. That is, with $\gamma = 0$, the agent is equally interested in its private goals and in the shared ones (and assumes the same for the partner); with $\gamma = 0$, the agent is solely interested in the shared goals (and also assumes the same for the partner).

This operation may seem quite artificial, especially as it implies a leap of faith on the part of the agent to effectively change its expectations about the partner's behavior (eq. 6). However, if we accept this assumption, we see that the task is made easier: in the general case, alignment reduces the agent-specific goal ambiguity, leading to better ability to focus and less positional ambiguity coming from the partner. Of course, one can construct examples where alignment does not help or even hurts; for instance, if both agents share all of their peaks, alignment not only will not help reduce ambiguity, but it can make the peaks sharper and hard to find. And as we will see, in the context of the system-level model, alignment becomes a natural capability.

---

[4] First-order ToM involves thinking about what some-one else is thinking or feeling; second-order ToM involves thinking about what someone is thinking or feeling about what someone else is thinking or feeling [95].

*2.11 System-level free energy*

Up until now, we have restricted ourselves to discussing our model at the level of individual agents and their local-scale interactions. We now take a higher vantage point and consider the implications of these local-scale interactions for global-scale system performance. We posit an ensemble of M identical copies of the two-agent subsystem above (i.e., 2*M*), each in its own independent environment, also assumed to be identical except for the position of the food source (see Figure 5).

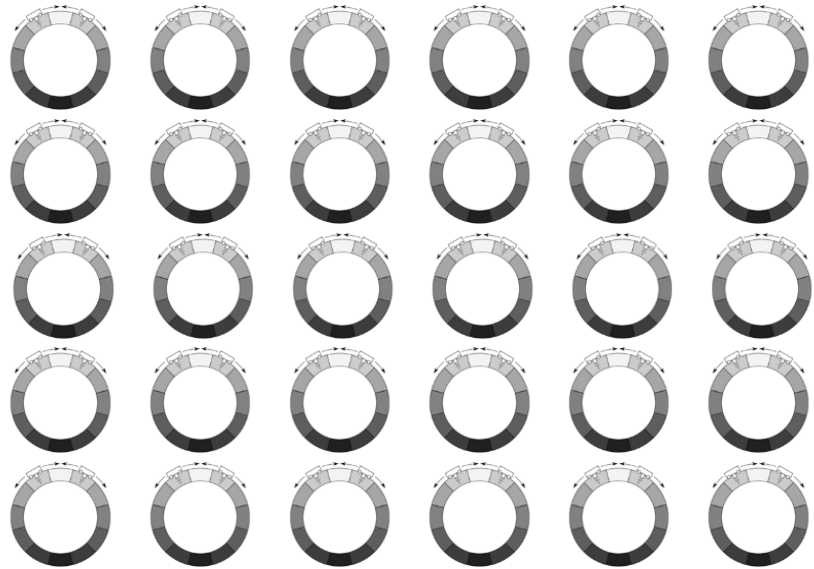

**Figure 5.** M identical copies of the two-agent subsystem

From this vantage point, each of the *2M* agents is now a "point particle", described only by its position $\psi^i$. More practically, we can change coordinates to $\varphi^i = \psi^i - \psi^i_0$; the tuple $b^{system} = (\varphi^i)_{i \in [1 \ldots 2M]}$ is then the set of internal states of the system as a whole. We can then define a system-level free energy as

$$F^{system} = D_{KL}(q^{empirical}(\varphi) \| p^{system}(\varphi)), \quad (7)$$

where $q^{empirical}(\varphi)_\lambda = \frac{1}{2M} \#\{\varphi^i \mid \varphi^i = \lambda\}$, the system's "variational density", is simply the empirical distribution of the various agents across all possible positions relative to their respective local food sources.

Note that the system is considered in isolation, and therefore the active states of the system are empty, $s^{system} = a^{system} = \emptyset$. Thus, if we are to analyze it as a single agent, the only thing it can do to minimize its free energy is to rearrange its internal states $\varphi^i$. This is analogous to a single time step in the simulation loop described at the individual level in the model pseudo code (see Figure 3). We can say that the 2*M* agents' aggregate behavior over an *entire run* of the model at the individual level implements a *single step* of approximate Bayesian inference for the system as a whole.

This in turn motivates defining the system's "generative density" as $p^{system}(\psi_0 \mid \psi) \sim N(\psi_0, \sigma)$: at any given point in time, the system "expects" the agent positions and food positions to be suitably close to each other. Thus, we can see that, to the extent that the agents move closer to their food positions, the system performs approximate Bayesian inference over this generative density, and we can evaluate the degree to which this inference is effective, by evaluating whether, and how quickly, $F^{system}$ is minimized. We return to the topic of system-level (active) inference in the discussion.

*2.12 Simulations*

We have thus defined this system at two altitudes, enabling us to perform simulations at the agent level and analyze their implied performance at the system level (as measured by system-level free energy). We can now use this framework to analyze the extent to which the two novel agent-level cognitive capabilities we introduced ("Theory of Mind" and "Goal Alignment") increase the system's ability to perform approximate inference at local and global scales. To explore the effects of agent-level cognitive capabilities on collective performance, we create four experimental conditions according to a 2x2 (Theory of Mind x Goal Alignment) matrix: Model 1 (Baseline), Model 2 (Theory of Mind), Model 3 (Goal Alignment), and Model 4 (Theory of Mind and Goal Alignment; see Table 1).

**Table 1**. 2x2 (Theory of Mind x Goal Alignment) permutations of our model

|  | -Theory of Mind | +Theory of Mind |
|---|---|---|
| **-Goal Alignment** | Model 1 (Baseline) | Model 2 (Theory of Mind, No Goal Alignment) |
| **+Goal Alignment** | Model 3 (Goal Alignment, No ToM) | Model 4 (Theory of Mind x Goal Alignment) |

Throughout, we use the same two agents, Agent A and Agent B. To establish meaningful variation in agent performance at the individual-scale, we parameterize an agent's perceptiveness to the physical environment (i.e., to the reliability of the information derived from its "chemical sensors"), by assigning one agent with "strong" perceptiveness (Agent A - Strong;) and the other agent with "weak" perceptiveness (Agent B - Weak).

We assign each agent with two targets, one shared (Shared Target) and one unshared (individual target or Target A and Target B). Accordingly, we assume each agent's desire distributions have both a shared peak (corresponding to a Shared Target) and an unshared peak (corresponding to Target A or Target B). Throughout, we measure both the collective performance (system-level free energy), as well as individual performance (distance from their closest target). In addition, we also capture their end-state desire distribution.

We implement simulations in Python (V3.7) using Google Colab (V1.0.0). As noted above, our implementation draws upon and extends an existing AIF model implementation developed in Python (V2.7) by van Shaik [66]. To ensure that the agent behavior is not an artefact of their specific location in the environment, we run 180 runs for each simulation for each experimental condition by randomizing their starting locations throughout the environment. The environment size was held constant at 60 cells.

*2.13 Model parameters*

Our four models were created by setting physical perceptiveness for the strong and weak agent and varying their ability to exhibit social perceptiveness and align goals. The parameter settings are summarized at the individual agent level as follows (see Figure 6 and Table 2):

- **Model 1** contains a self-actualization loop driven by physical perceptiveness. Physical perceptiveness (individual skill parameter; range [.01,.99]) is varied such that Agent A is endowed with strong perceptiveness (.99) and Agent B is endowed with weak perceptiveness (.05).

- **Model 2** is made up of a self-actualization loop and a partner-actualization loop (instantiating ToM). The other-actualization loop is implemented by setting the value of alterity (ToM or social perceptiveness parameter; range [0.01,.99]) as .20 for the weak agent and 0 for the strong agent. This parameterization helps the weak agent use social information to navigate the physical environment. These two loops implement a single (non-separable) free energy functional: The weak agent's inferences from their stronger partner's behavior serve to refine its beliefs about its position in the environment.

- **Model 3** entails a self-actualization loop (but no partner-actualization loop) as well as enforces the pursuit of a common goal (set alignment = 1) by fully suppressing their unshared goals (alignment parameter; range [0,1]). In this simplified implementation, we assume that goal alignment is a relational/dyadic property such that both partners exhibit the same level of alignment towards each other. This is akin to partners fully exploring each other's targets and agreeing to pursue their common goal. Setting alignment lower the 1 will increase the relative weighting of unshared goals and cause them to compete with their shared goals

- **Model 4** includes both cognitive features: self- and partner-actualization loops for the weak agent (instantiating ToM; alterity = 0.2) and complete goal alignment between agents

**Table 2.** Parameterization of agent abilities Models 1- 4

|  | **Model 1** Baseline | | **Model 2** Theory of Mind | | **Model 3** Goal Alignment | | **Model 4** ToM x Goal Alignment | |
| --- | --- | --- | --- | --- | --- | --- | --- | --- |
| Parameter | Agent B | Agent B | Agent A | Agent B | Agent A | Agent B | Agent A | Agent B |
| Physical perceptiveness (.01, .99) | **.99** | **.05** | .99 | .05 | .99 | .05 | .99 | .05 |
| Alterity, $\alpha$ (.01, .99) | .00 | .00 | **.00** | **.20*** | .00 | .00 | **.00** | **.20*** |
| Goal Alignment, $\gamma$ (0, 1) | 0 | 0 | 0 | 0 | 1 | 1 | 1 | 1 |

*Alternative results for simulations with alterity set at $\alpha = .5$ exhibit a similar pattern of results for Model 2 and Model 4.

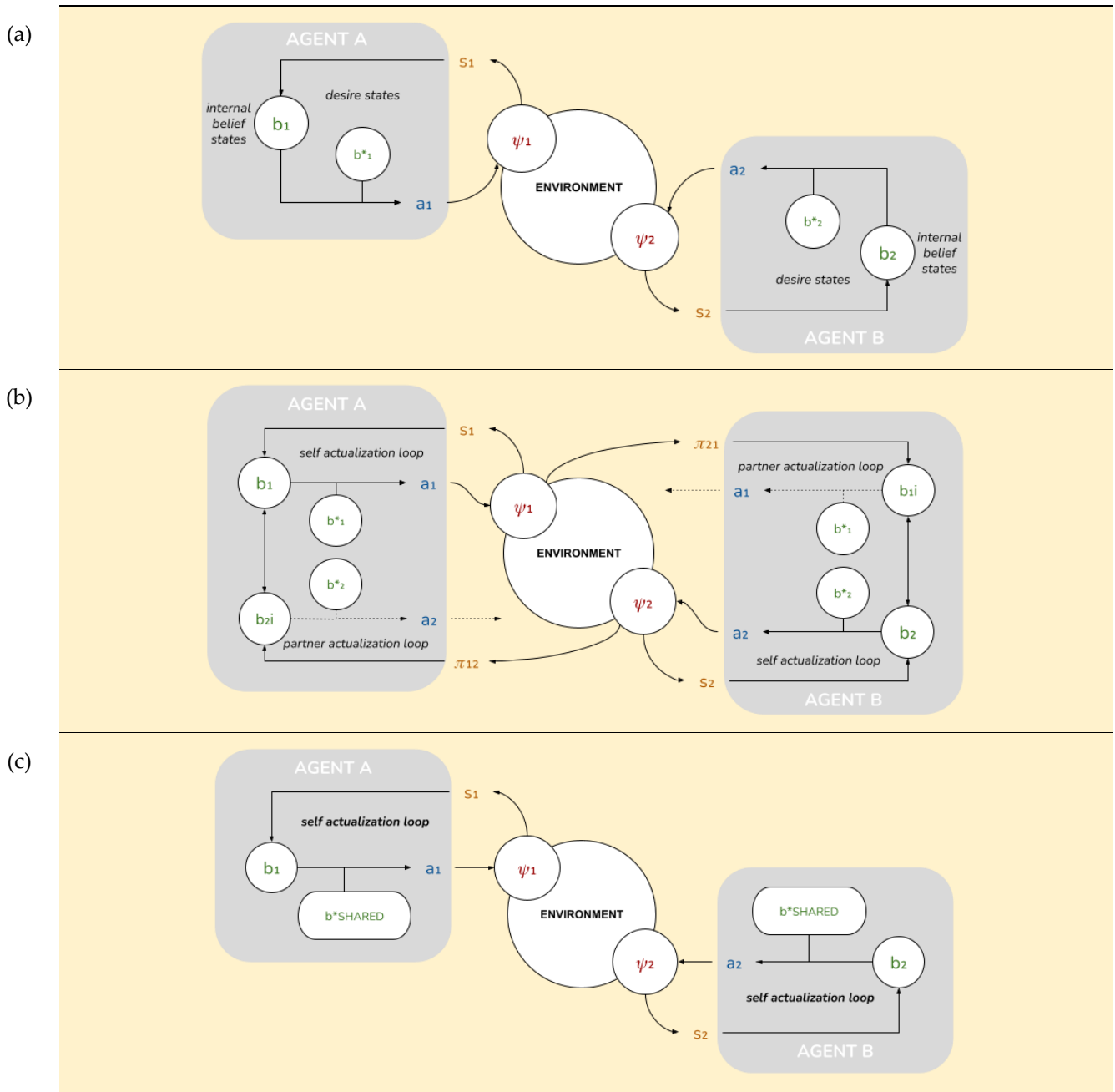

(d)

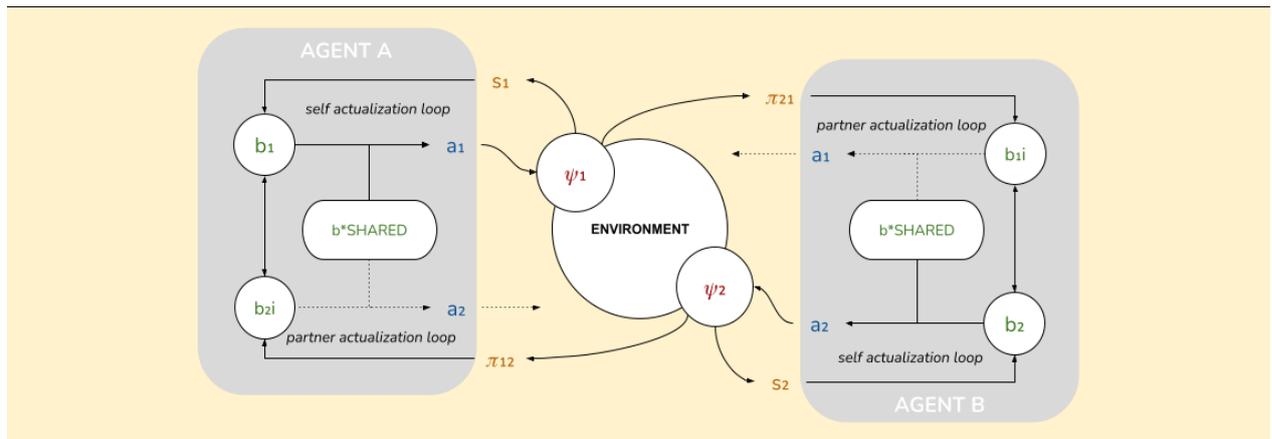

**Figure 6.** Models. (a) Model 1 - Baseline; (b) Model 2 – Theory of Mind; (c) Model 3 – Goal Alignment; (d) Model 4 – Theory of Mind with Goal Alignment

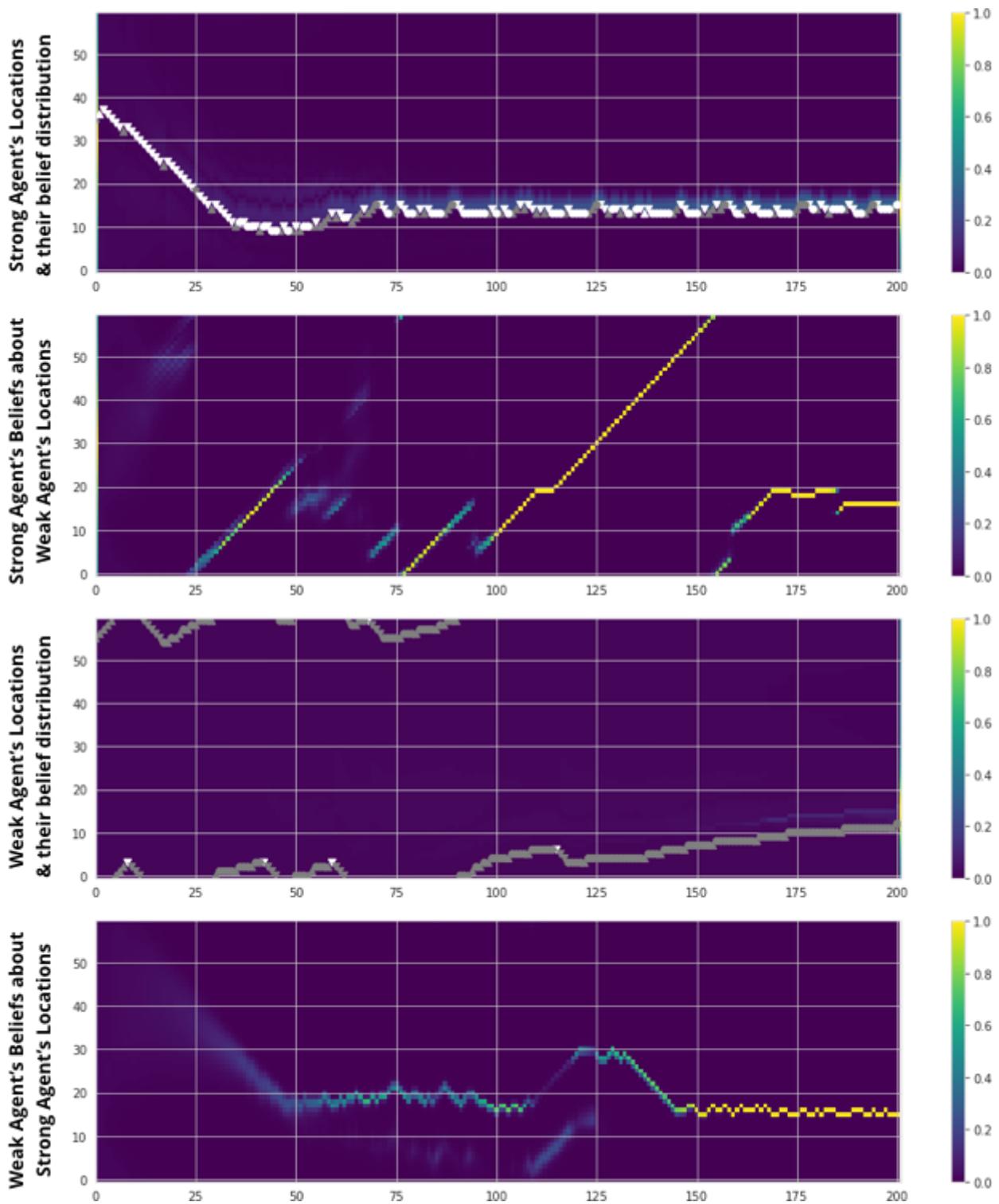

**Figure 7.** Results from a single run of Model 4 over 200 epochs. Agents' Shared Target position is set at location 15. Actual agent positions are illustrated as single dots for each epoch on the top graph, colored white when s=1 and gray when s=0. The background of the top graphs plots the agents' belief distribution of their own position, from dark blue (0) to bright yellow (1). The bottom graphs plot the agents' belief distribution of their partner's position, on the same scale.

## 3. Results

*3.1 Illustration of Agent-level Behavior*

In Figure 7, we show typical results from a single run of a single two-agent subsystem (Model 4: ToM with Goal Alignment) to illustrate qualitatively how the two cognitive capabilities introduced enable agent-level performance. In this example, Goal Alignment enters the picture at the outset; although each agent has two targets, they both only ever pursue their shared target.

The evolution of the two agents' behavior and beliefs over this run demonstrates the key features of interplay between sensory and partner inputs, and how ToM moderates the influence of partner inputs on an agent's behavior. Using its high perceptiveness, A identifies its own position around epoch 25-50, and quickly thereafter, directs itself towards the food position and remains stable there (top left). Meanwhile, for most of the run, B has no strong sense of its own position, and therefore its movement is highly random and undirected; at around epoch 150, it finally starts exhibiting a sharper (light blue) belief and converging to the target (top right). This is the same moment when B is finally able to disambiguate A's behavior (from green to yellow), which, via ToM, enables B's belief to become sharper (bottom right). Meanwhile, A can't make sense of B's random actions: the partner distribution it infers is unstable. But because A has ToM = 0, it doesn't take any of these misleading cues into account when deciding its own beliefs (bottom left).

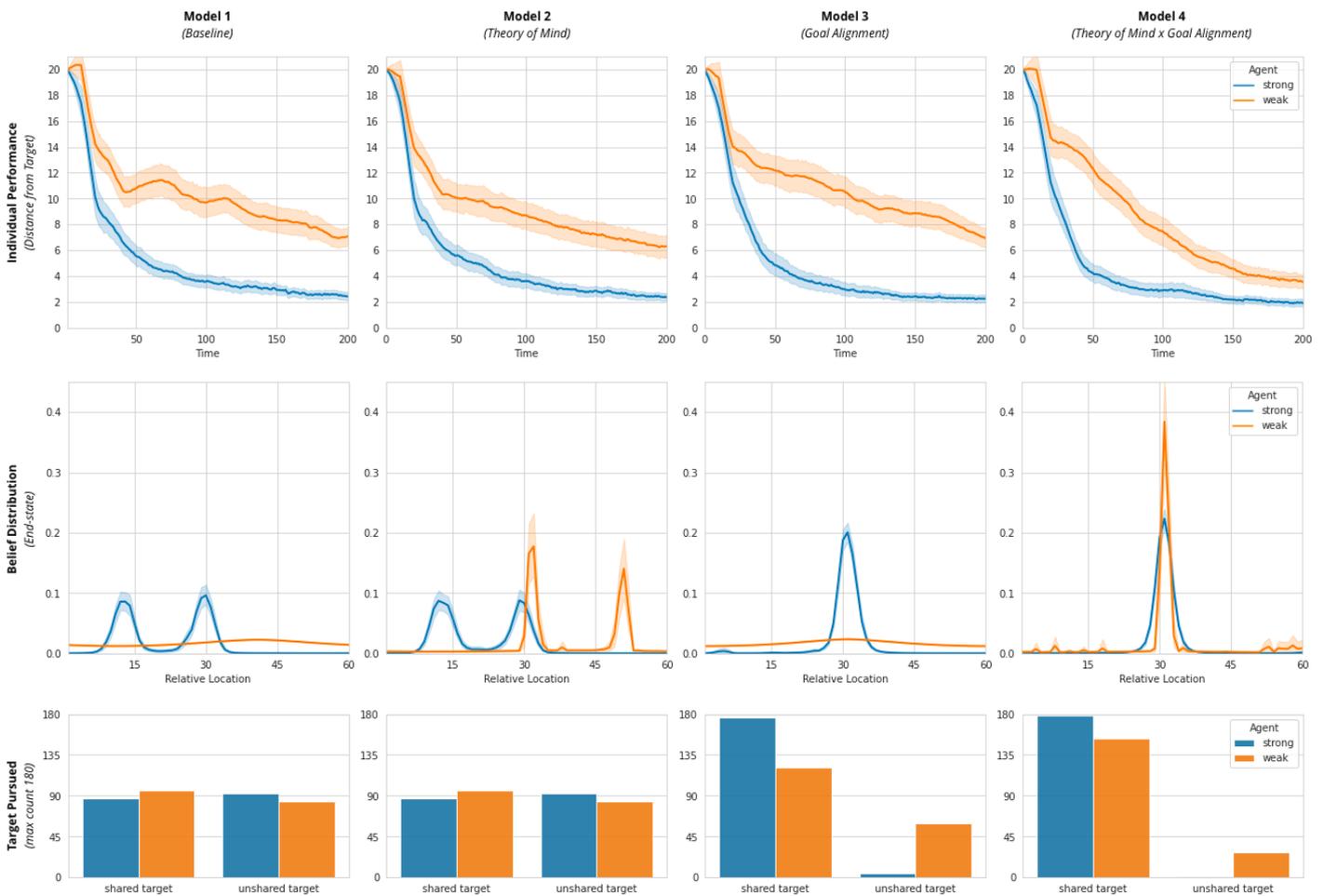

**Figure 8.** Simulation results of Agent A (strong; blue) and Agent B (weak; orange) in all four models. Row 1: Individual performance as time taken to reach a target position. Row 2: End state belief distribution of target location (Shared Target = 30; A's Target = 15; B's Target = 45). Row 3: Distribution of targets pursued in 180 runs.

*3.2 Simulation Results*

Model 1 lends face validity to the two-agent simulation setup. Figure 8 (Row 1, Model 1) demonstrates that, on average, the strong agent (endowed with high physical perceptiveness) converges to an end-state belief faster more accurately (closer to one of their individual targets) than the weak agent with severely diminished physical perceptiveness. This difference in individual performance can be attributed to the stark difference in agents' ability to form strong beliefs about the location of their target (see Figure 8: Row 2, Model 1). Agents show no clear preference for either shared or unshared targets (Figure 8: Row 3, Model 1).

In model 2, the weak agent possesses 'Theory of Mind'. This allows it to infer information about their own location in the environment by observing their partner's actions. This is evidenced by the emergence of two-sharp peaks in the weak agent's end-state belief distribution (Figure 8: Row 2, Model 2). Consequently, we see an improvement in the weak agent's individual performance (the agent converges faster on an end-state belief faster than in Model 1). Collective performance (Figure 9: System's free energy) does not appear to improve between Model 1 and Model 2. This may be because agents solely focus on achieving their individual goals (and do not understand any distinction between individual and system level goals). This is evidenced by the fact that of the 180 simulation runs each of Model 1 and Model 2, both agents end up pursuing their shared and unshared targets with roughly equal probability (Figure 8: Row 3, Model 1 and 2) .

In Model 3, when both agents possess an ability for Goal Alignment, but the weak agent does not have the benefit of Theory of Mind, we see that both agents are biased towards pursuing the shared system goal (Figure 8: Row 3, Model 3). Accordingly, at the system level we see naturally higher collective performance—Model 3 clearly has lower system-level free energy compared to both Model 1 and Model 2 (see Figure 9). At the individual-level, however, the weak agent performs worse on average than it did in Model 2 and converges more slowly towards its goals (Figure 8: Row 1, Model 3). It appears that Goal Alignment helps improve system performance by reducing the ambiguity of multiple possible targets, but Goal Alignment does not help the weak agent compensate for low physical perceptiveness.

Finally, as expected, in Model 4, which combines Theory of Mind and Goal Alignment, we see a clear improvement in both individual and collective performance (Figure 8: Row 1, Model 4 and Figure 9: Model 4, respectively). The combination of Theory of Mind (for the weak agent) and Goal Alignment (for both agents) appears to enable the weak agent to overcome its poor physical perceptiveness and converge on a single unambiguous end-state belief. This achievement is illustrated by the sharp and overlapping single-peaked end-state belief structure achieved by both agents in model 4 (Figure 8: Row 2, Model 4). This model suggests that collective performance is highest when individual agents' individual states align with the global system state.

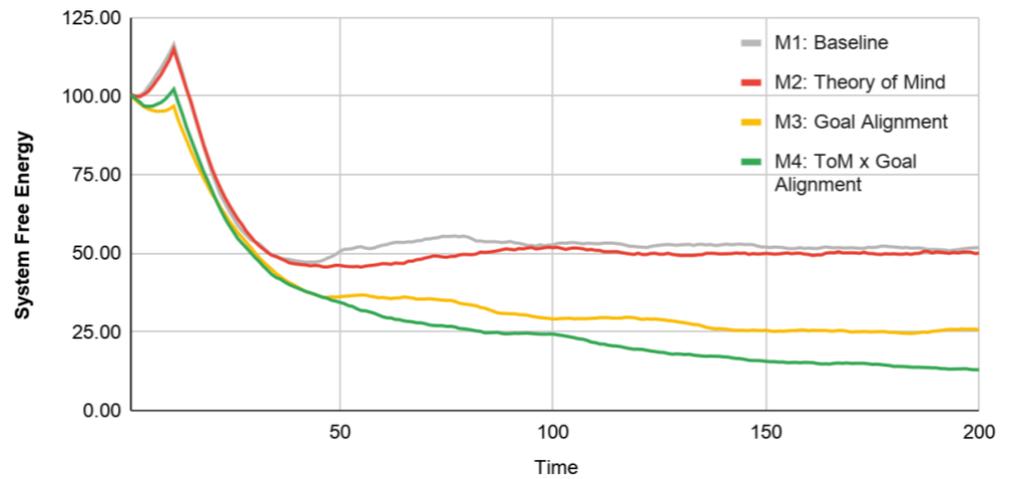

**Figure 9.** Collective performance plotted as the system's free energy. Lower free energy denotes higher system performance. Gradient descent on free energy suggests a system that performs (active) Bayesian inference.

## 4. Discussion

A formal understanding of collective intelligence in complex adaptive systems requires a formal description, within a single multiscale framework, of how the behavior of a composite system and its subsystem components co-inform each other to produce behavior that cannot be explained at any single scale of analysis. In this paper we make a contribution toward this type of formal grasp of collective intelligence, by using AIF to posit a computational model that connects individual-level constraints and capabilities to collective-level behavior. Specifically, we provide an explicit, fully specified two-scale system where free energy minimization occurs at both scales, and where the aggregate behavior of agents at the faster/smaller scale can be rigorously identified with the belief-optimization (a.k.a. "inference") step at the slower/bigger scale. We introduce social cognitive capabilities at the agent level (Theory of Mind and Goal Alignment), which we implement directly through AIF. Further, illustrative results of this novel approach suggest that such capabilities of individual agents are directly associated with improvements in the system's ability to perform approximate Bayesian inference or minimize variational free energy. Significantly, improvements in global-scale inference are greatest when local-scale performance optima of individuals align with the system's global expected state (e.g., Model 4). Crucially, all of this occurs "bottom-up", in the sense that our model does not provide exogenous constraints or incentives for agents to behave in any specific way; the system-level inference emerges as a product of self-organizing AIF agents endowed with simple social cognitive mechanisms. The operation of these mechanisms improves agent-level outcomes by enhancing agents' ability to minimize free energy in an environment populated by other agents like it.

Of course, our account does not preclude or dismiss the operation of "top-down" dynamics, or the use of exogenous incentives or constraints to engineer specific types of individual and collective behavior. Rather, our approach provides a principled and mechanistic account of bio-cognitive systems in which "bottom-up" and "top-down" mechanisms may meaningfully interplay to inform accounts of behavior such as collective intelligence [4]. Our results suggest that models such as these may help establish a mechanistic understanding of how collective intelligence evolves and operates in real-life systems, and provides a plausible lower bound for the kind of agent-level cognitive capabilities that are required to successfully implement collective intelligence in such systems.

*4.1 We demonstrate AIF as a viable mathematical framework for modelling collective intelligence as a multiscale phenomenon*

This work demonstrates viability of AIF as a mathematical language that can integrate across scales of a composite bio-cognitive system to predict behavior. Existing multiscale formulations of AIF [29,30], while more immediately useful for understanding the behavior of docile subsystem components like cells in a multicellular organism or neurons in the brain, do not yet offer clear predictions about the behavior of collectives composed of highly autonomous AIF agents that engage in reciprocal self-evidencing with each other as well as with the physical (non-social) environment [34]. What's more, existing toy simulations of multiscale AIF engineer collective behavior as a predestination—either as a prior in an agent's generative model [33], or by default of an environment that consists solely of other agents [48,49]. We build upon these accounts by using AIF to first posit the minimal information-theoretical patterns (or "adaptive priors"; see [32]) that would likely emerge at the level of the individual agent to allow that agent to persist and flourish in an environment populated by other AIF agents [53]. We then examine the relationship between these local-scale patterns and collective behavior as a process of Bayesian inference across multiple scales. Our models show that collective intelligence can emerge endogenously in a simple goal-directed task from interaction between agents endowed with suitably sophisticated cognitive abilities (and without the need for exogenous manipulation or incentivization).

Key to our proposal is the suggestion that collective intelligence can be understood as a dynamical process of (active) inference at the global-scale of a composite system. We operationalize self-organization of the collective as a process of free energy minimization or inference based on sensory (but not active) states (for a previous attempt to operationalize collective behavior as both active and sensory inference, see [65]). In a series of four models, we demonstrate the responsiveness of this system-level measure to learning effects over time; the progression of each Model exhibits a pattern akin to a gradient descent on free energy, evoking the notion that a system that performs (active) Bayesian inference. Further, stepwise increases in cognitive sophistication at the individual level show a clear reduction in free energy, particularly between Model 1 (Baseline) and Model 4 (Theory of Mind x Goal Alignment). These illustrative results establish a formal, causal link between behavioral processes across multiple scales of a complex adaptive system.

Going further, we can imagine an extension of this model where the collective system interacts with a non-trivial environment, but at a slower time scale, such that a complete simulation run of all 2M agents corresponds to a single belief optimization step for the whole system, after which it acts on the environment and receives sensory information from it (manifested, for example, as changes in the agents' food sources). In this extended model (see Figure 10), and if the agent-specific parameters (alterity/Theory of Mind ($\alpha$), and Goal Alignment ($\gamma$)) could be made endogenous (either via selective mechanisms via some other learning mechanisms; see [39,67]) we would expect to see the system finding (non-zero) values of these parameters that optimize its free energy minimization. For example, it is likely that a system would select for higher values of $\gamma$ (Goal Alignment) when both agents' end-state beliefs and actual target locations mutually cohere, or higher values of $\alpha$ for agents with weaker perceptiveness. Interestingly, this would show that degrees of Theory of Mind and Goal Alignment are capabilities that would be selected for or boosted at these longer time scales, providing empirical support for the heuristic arguments made for their existence in our model and in human collective intelligence research more generally [4].

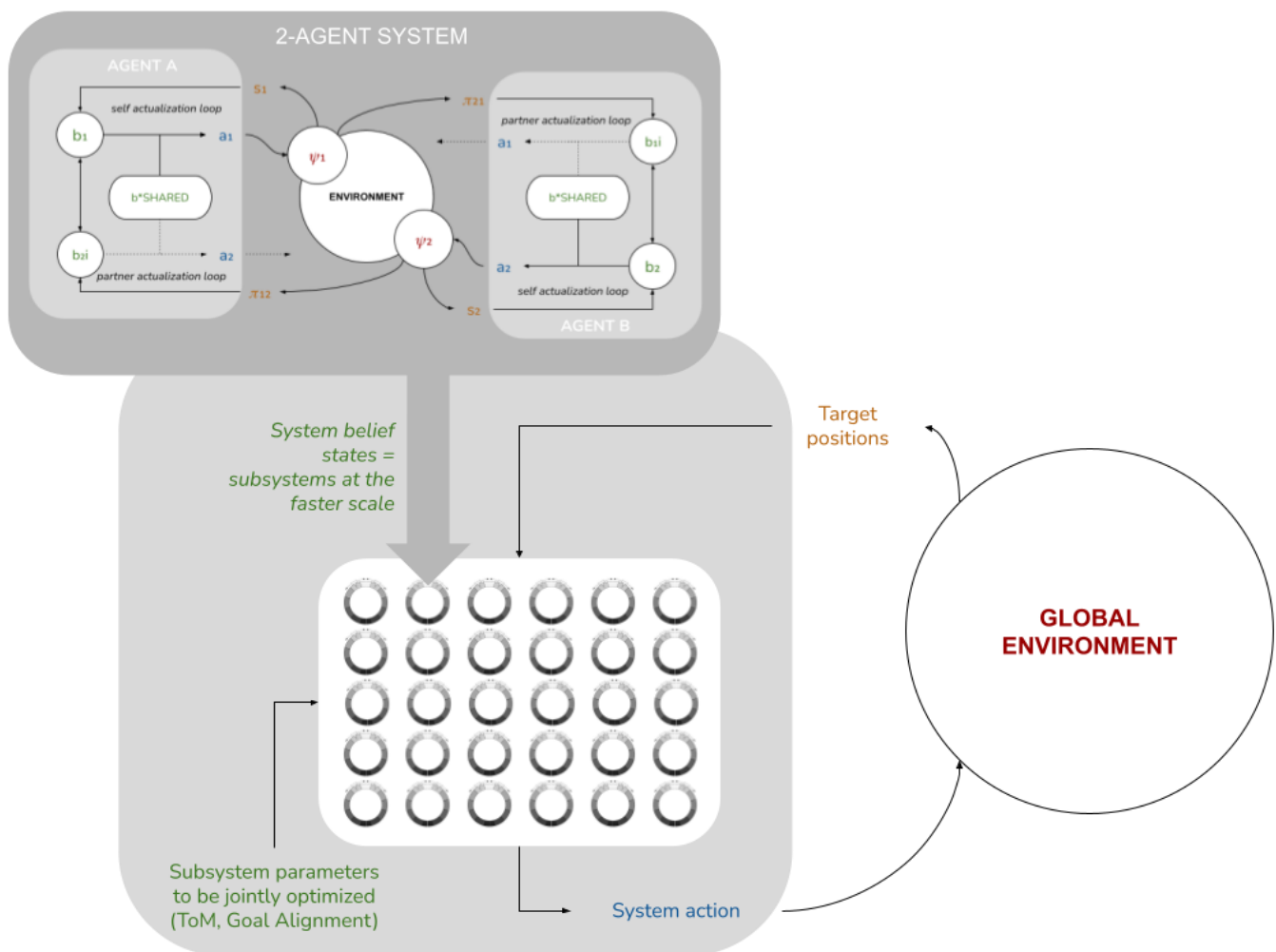

**Figure 10.** A notional complete two-scale model where agent-specific parameters are endogenized.

*4.2 AIF sheds light on dynamical operation of mechanisms that underwrite collective intelligence*

In this way, AIF offers a paradigm through which to move beyond the methodological constraints associated with experimental analyses of the relationship between local interactions and collective behavior [12]. Even our very rudimentary 2-Agent AIF model proposed here offers insight into the dynamic operation and function of individual cognitive mechanisms for individual and collective level behavior. In distinct contrast to laboratory paradigms that usually rely on low-dimensional behavioral "snapshots" or summaries of behavior to verify linearly causal predictions about individual and collective phenomena, our computational model can be used to explore the effects of fine-grained, agent- and collective-level variations in cognitive ability on individual and collective behavior in real time.

For example, by parameterizing key cognitive abilities (Theory of Mind and Goal Alignment), our model shows that it is not necessarily a case of "more is better" when it comes to cognitive mechanisms underlying adaptive social behavior and collective intelligence. If an agent's level of social perceptiveness (Theory of Mind) were too low, it is likely that agents would miss vital performance-relevant information about the environment populated by other agents; if an agent's Theory of Mind were too high, it may instead over-index on partner belief states as an affordance for own beliefs (a scenario of "blind leading the blind"). We show that canonical cognitive abilities such as Theory of Mind and Goal Alignment can function across multiple scales to stabilize and reduce the computational uncertainty of an environment made up of other AIF agents, but only when these abilities are optimally tuned to a "goldilocks" level that is suitable to performance in that specific environment.

The essence of this proposal is captured by empirical research of attentional processes of human agents that engage in sophisticated joint action [68,69]. For instance, athletes in novice basketball teams are found to devote more attentional resources to tracking and monitoring their own teammates, while expert teams spend less time attending to each other and more time instead attending to the socio-technical task environment [70]. Viewed from the perspective of AIF, in both novice and expert teams, agents likely differentially deploy physical and social perceptiveness at levels that make sense for pursuing collective performance in a given situation; novices may stand to gain more from attending to (and therefore learning from) their teammates (recall our Agent B in Model 2 who leverages Theory of Mind to overcome weak physical perceptiveness, for example); while experts might stand to gain more from down-regulating social perceptiveness and redirecting limited attentional resources to physical perception of the task or (adversarial) social environment [71,72].

To circle back to the specific domain of human collective intelligence examined in organizational psychology and management, it seems likely that social perceptiveness may indeed be an important factor (among many) that underwrites collective intelligence. But this may be especially the case in the context of unacquainted teams of "WEIRD" experimental subjects [73] who coordinate for a limited number of hours in a contrived laboratory setting [3]. If the experimental task were to be translated to a real-world performance setting (e.g. one involving high-stakes or elite performance requirements), or if that same team of experimental subjects were to persist over time beyond the lab in a randomly fluctuating environment, it is conceivable that a premium for social perceptiveness may give way to demands for other types of abilities needed to continue to gain performance-relevant information from the task environment (e.g., through physical perceptiveness of the task environment). Viewed from this perspective, the true "special sauce" of collective intelligence (and individual intelligence, for that matter; see [74]) may turn out not to be one or other discrete or reified individual or team level ability per se (e.g., social perceptiveness), but instead a collective ability to nimbly adjust the volumes of multiple parameters to foster specific information-theoretic patterns conducive to minimizing free energy across multiple scales and over specific, performance-relevant time periods.

In this spirit, the computational approach we adopt here under AIF affords a dynamical and situational perspective on team performance that may offer important insights into long-standing and nascent hypotheses concerning the causal mechanisms of collective intelligence. For instance, our model is well positioned to investigate the long-proposed (but hitherto unsubstantiated) claim that successful team performance, and by extension, collective intelligence, depends on balancing a tradeoff between cognitive diversity and cognitive efficiency [4] (p. 421). Likewise, our approach could help elucidate mechanisms and dynamics through which memory, attention, and reasoning capabilities become distributed through a collective, and the conditions in which these "transactive" processes [75] facilitate emergence of intelligent behavior [70,76,77]. In either case, our model would simply require specification with the appropriate individual-level cognitive abilities or priors. For example, to better understand the causal relationship between transactive knowledge systems and collective intelligence, our model could leverage recent empirical research that observes a connection between individual agents' metacognitive abilities (e.g., perception of others' skills, focus, and goals), the formation of transactive knowledge systems, and a collective's ability to adapt to a changing task environment [78]. On an important and related note to these opportunities for future research, efforts to simulate human collective intelligence should strive to develop models composed of two or more agents to better mimic human-like coordination dynamics [43,79].

*4.3 Increases in system performance correspond with alignment between an agent's local and global optima.*

A key insight from our models, and worthy of further investigation, is that the greatest improvement in collective intelligence (Model 4; measured by global-scale inference) occurs when local-scale performance optima of individuals align with the system's global expected state. This effect can be understood as individuals jointly implementing approximate Bayesian inference of the system's expectations. In effect, our model suggests that multi-scale alignment between lower- and higher-order states may contribute to the emergence of collective intelligence.

Alignment between local and global states might sound like an obvious prerequisite for collective intelligence, particularly for more docile AIF agents such as neurons or cells (it is near impossible to imagine a scenario in which a neuron or cell could meaningfully persist without being spatially aligned with a superordinate agent; see Palacios et al., 2020). But our model exemplifies a more subtle form of alignment, based on a loose coupling between scales through a system's generative model (section 2.11), enabling the extension of this idea to scenarios where the local and global optimizations may be taking place in arbitrarily distinct and abstract state spaces [42,44]. By now it is well understood that coordinated human behavior relies for its stability and efficacy on an intricate web of biologically evolved physiological and cognitive mechanisms [80,81], as well as culturally evolved affordances of language, norms, and institutions [82]. But precisely how these various mechanisms and affordances—particularly those that are separated across

scales—coordinate in real or evolutionary time to enable human collective phenomena remains poorly understood [29,83,84].

Computational models capable of formally representing multiscale alignment may help reorganize and clarify causal relationships between the various hypothesized physiological, cognitive, and cultural mechanisms hypothesized to underpin human collective behavior [85]. For example, a computational model such as the one proposed here could conceivably be adapted to help more systematically test the burgeoning hypothesis that coordination between basal physiological, metabolic and homeostatic processes at one scale of organization and linguistically mediated processes of interaction and exchange at another scale determine fundamental dynamics of individual and collective behavior [86–88].

Future research should aspire to examine causal connections between a fuller range of meaningful scales of behavior. In the case of human collectives, meaningful scales of behavior could extend from the basal mechanisms of physiological energy, movement, and emotional regulation on the micro scale [89,90], to linguistically- (and now digitally-) mediated social informational systems at the meso scale [91] to global socio-ecological systems at the macro scale [84,92,93]. As we have demonstrated here, the key requirement for the development of such multiscale models under AIF is faithful construction of the appropriate generative models at each scale to provide the mechanistic "missing links" between AIF and the phenomena to be explained—a task that will require tremendously innovative and intelligent collective behavior on the part of a diverse range of agents.

*The patterns that crop up again and again in successful space are there because they are in fundamental accord with characteristics of the human creature. They allow him to function as a human. They emphasize his essence—he is at once an individual and a member of a group. They deny neither his individuality nor his inclination to bond into teams. They let him be what he is.*

- DeMarco and Lister [94](1987, p.90)

**Author Contributions:** Conceptualization, R.K., J.T., and P.G.; methodology, R.K., P.G. and J.T; software, R.K., P.G. validation, R.K., P.G; formal analysis, R.K.; investigation, R.K., P.G. and J.T; resources, R.K., J.T., and P.G.; data curation, P.G. and R.K.; writing—original draft preparation, J.T. & R.K.; writing—review and editing, and J.T, R.K. and P.G.; visualization, P.G., R.K., J.T. All authors have read and agreed to the published version of the manuscript.

**Funding:** This research received no external funding

**Data Availability Statement:** Model code can be found and implemented via this link to Google Colab:

https://colab.research.google.com/drive/1CKdPTy8LD-Mpxc7kXy47m_fmCq44BT5u?usp=sharing

**Acknowledgments:**

**Conflicts of Interest:** The authors declare no conflict of interest